# XHIP: An Extended Hipparcos Compilation

*Erik Anderson [1], Charles Francis [2]*

[1] 360 Iowa Street, Ashland, OR 97520, USA (erik@astrostudio.org)

[2] 125 Elphinstone Road, Hastings, TN34 2EG, UK

**Abstract**

We present the Extended Hipparcos Compilation (XHIP), a database of all stars in the New Reduction of the Hipparcos Catalog extensively cross-referenced with data from a broad survey of presently available sources. The resulting collection uniquely assigns 116 096 spectral classifications, 46 392 radial velocities, and 18 549 homogenized iron abundances [Fe/H] to Hipparcos stars. Stellar classifications from SIMBAD, indications of multiplicity from CCDM or WDS, stellar ages from the Geneva-Copenhagen Survey III, supplemental photometry from 2MASS and SIMBAD, and identifications of exoplanet host stars are also included. Parameters for solar encounters and Galactic orbits are calculated for a kinematically complete subset. Kinetic bias is found to be minimal. Our compilation is available through the Centre de Données astronomiques de Strasbourg as Catalog V/137B.

## 1 Introduction

High-precision astrometry is vital to modern astronomical studies. The precision attained by ESA's Hipparcos space mission remains today the state-of-the-art. ESA's much-anticipated Gaia mission intends to supersede Hipparcos' astrometric precision by two orders of magnitude, but the projected release date for Gaia's final results is still a decade away (Lindegren, 2010). In the meantime, Hipparcos astrometry will continue to be widely used in both scientific and non-scientific applications.

It is often necessary to cross reference Hipparcos stars with other types of data. The Extended Hipparcos (XHIP) dataset is an up-to-date compilation of supplemental data in a single database. The methods of constructing the compilation are documented in §2. We identify members of open clusters and stellar associations in §3. Our tests for kinetic bias are reported in §4. We document our metallicity calibration procedure in §5. We give our conclusions in §6.

## 2 Construction

The contents of the XHIP compilation is summarized Table 1. The basis for our compilation is the *Hipparcos, the New Reduction of the Raw Data* ("HIP2", van Leeuwen, 2007), which largely improved the astrometry over the original Hipparcos catalog ("HIP1", ESA, 1997). Many non-stellar objects cataloged in HIP1 were also removed in HIP2 (though some entries identified by SIMBAD as planetary nebulae and one quasar do remain). The 117 955 entries in HIP2 constitute the master list common to *main.dat, photo.dat,* and *biblio.dat* – described in detail below.

Left section (main.dat):

| main.dat | Label | Units | Explanation |
|---|---|---|---|
| 1 | HIP | --- | Hipparcos identifier |
| 2 | Comp | --- | Component(s) |
| 3 | Classes | --- | SIMBAD classifications |
| 4 | Groups | --- | Cluster or assn. memberships |
| 5 | RAdeg | deg | Right ascension |
| 6 | DEdeg | deg | Declination |
| 7 | Plx | mas | Trigonometric parallax |
| 8 | pmRA | mas/yr | Proper motion in RA |
| 9 | pmDE | mas/yr | Proper motion in DE |
| 10 | e_RA | mas | Standard error on RA |
| 11 | e_DE | mas | Standard error on DE |
| 12 | e_Plx | mas | Standard error on Plx |
| 13 | e_pmRA | mas/yr | Standard error on pmRA |
| 14 | e_pmDE | mas/yr | Standard error on pmDE |
| 15 | r_Plx | --- | Reference for parallax |
| 16 | r_pm | --- | Reference for proper motion |
| 17 | GLON | deg | Galactic longitude |
| 18 | GLAT | deg | Galactic latitude |
| 19 | Dist | pc | Heliocentric distance |
| 20 | e%_Dist | --- | Distance error as percentage |
| 21 | pmGLON | mas/yr | Proper motion in GLON |
| 22 | pmGLAT | mas/yr | Proper motion in GLAT |
| 23 | X | pc | Distance component X |
| 24 | Y | pc | Distance component Y |
| 25 | Z | pc | Distance component Z |
| 26 | RGal | pc | Distance from Galactic center |
| 27 | vT | km/s | Transverse velocity |
| 28 | SpType | --- | Spectral type (MK, HD, etc.) |
| 29 | Tc | --- | Temperature class code |
| 30 | Lc | --- | Luminosity class code |
| 31 | RV | km/s | Radial velocity |
| 32 | e_RV | km/s | Standard error on RV |
| 33 | q_RV | --- | Quality flag on RV |
| 34 | [Fe/H] | dex | Iron abundance |
| 35 | e_[Fe/H] | dex | Standard error on [Fe/H] |
| 36 | q_[Fe/H] | --- | Quality flag on [Fe/H] |
| 37 | age | Gyr | Age, in billions of years |
| 38 | clage | Gyr | Lower confidence limit on age |
| 39 | chage | Gyr | Upper confidence limit on age |
| 40 | U | km/s | Velocity component U |
| 41 | V | km/s | Velocity component V |
| 42 | W | km/s | Velocity component W |
| 43 | UVW | km/s | Total heliocentric velocity |
| 44 | Dmin | pc | Distance of solar encounter |
| 45 | Tmin | kyr | Timing of Dmin |
| 46 | ecc | --- | Total eccentricity |
| 47 | phi | deg | Pericenter position angle |
| 48 | a | pc | Semi-major axis |
| 49 | b | pc | Semi-minor axis |
| 50 | c | pc | Focus-to-center distance |
| 51 | L | pc | Semilatus rectum |
| 52 | Rmin | pc | Orbital radius at pericenter |
| 53 | Rmax | pc | Orbital radius at apocenter |
| 54 | planets | --- | Number of exoplanets |
| 55 | methods | --- | Planet discovery method(s) |

Right section (photo.dat):

| photo.dat | Label | Units | Explanation |
|---|---|---|---|
| 1 | HIP | --- | Hipparcos identifier |
| 2 | Hpmag | mag | Median mag. in Hip. system |
| 3 | e_Hpmag | mag | Standard error on Hpmag |
| 4 | m_Hpmag | mag | Reference flag for Hpmag |
| 5 | Hpmax | mag | Hpmag at maximum |
| 6 | Hpmin | mag | Hpmag at minimum |
| 7 | Per | days | Variability period |
| 8 | HvarType | --- | Variability type |
| 9 | Umag | mag | Magnitude in Johnson U |
| 10 | Bmag | mag | Magnitude in Johnson B |
| 11 | Vmag | mag | Magnitude in Johnson V |
| 12 | Rmag | mag | Magnitude in R |
| 13 | Imag | mag | Magnitude in I |
| 14 | Jmag | mag | J selected default magnitude |
| 15 | Hmag | mag | H selected default magnitude |
| 16 | Kmag | mag | K selected default magnitude |
| 17 | e_Jmag | mag | J total magnitude uncertainty |
| 18 | e_Hmag | mag | H total magnitude uncertainty |
| 19 | e_Kmag | mag | K total magnitude uncertainty |
| 20 | 2MASS | --- | 2MASS source designation |
| 21 | q_JHK | --- | JHK photometric quality flag |
| 22 | B-V | mag | Johnson B-V color |
| 23 | V-I | mag | Johnson V-I color |
| 24 | e_B-V | mag | Standard error on B-V |
| 25 | e_V-I | mag | Standard error on V-I |
| 26 | Hp_Mag | mag | Absolute Hpmag |
| 27 | U_Mag | mag | Absolute Magnitude U |
| 28 | B_Mag | mag | Absolute Magnitude B |
| 29 | V_Mag | mag | Absolute Magnitude V |
| 30 | R_Mag | mag | Absolute Magnitude R |
| 31 | I_Mag | mag | Absolute Magnitude I |
| 32 | J_Mag | mag | Absolute Magnitude J |
| 33 | H_Mag | mag | Absolute Magnitude H |
| 34 | K_Mag | mag | Absolute Magnitude K |
| 35 | Lum | Lsun | Stellar luminosity |
| 36 | magmin | mag | Magnitude V at Tmin |

biblio.dat:

| biblio.dat | Label | Units | Explanation |
|---|---|---|---|
| 1 | HIP | --- | Hipparcos identifier |
| 2 | HD | --- | Henry Draper cat. identifier |
| 3 | Con | --- | Constellation membership |
| 4 | Atlas | --- | Millennium Star Atlas pg. # |
| 5 | Coords | --- | RA, DE in compact format |
| 6 | Name | --- | Star Name(s) |
| 7 | GrpNames | --- | Cluster/Assn. Names |
| 8 | r_Comp | --- | Reference for Comp |
| 9 | r_SpType | --- | Reference for SpType |
| 10 | r_RV | --- | Reference for RV |
| 11 | r_[Fe/H] | --- | Reference for [Fe/H] |

keycodes.dat:

| keycodes.dat | Label | Units | Explanation |
|---|---|---|---|
| 1 | KeyCode | --- | Code from r_RV or r_[Fe/H] |
| 2 | BibCode | --- | Bibliographic code |

groups.dat:

| groups.dat | Label | Units | Explanation |
|---|---|---|---|
| 1-36 | | --- | [See readme file] |

**Table 1:** Contents of the XHIP dataset. The main body of information is found in *main.dat*, photometric data is kept in *photo.dat*, bibliographic references are listed in *biblio.dat*, the bibliographic key for radial velocity and iron-abundance sources is in *keycodes.dat*, and statistics for star clusters and associations are provided in *groups.dat*.

## 2.1  Comp, Classes, & Groups (*main.dat*)

Component designations of multiple star systems are provided in the *Comp* column. They are sourced from the *Catalog of Components of Double & Multiple Stars* ("CCDM", Dommanget & Nys, 2002) and *The Washington Visual Double Star Catalog, version 2010-11-21* ("WDS", Mason et al., 2001-2010) in that order of preference. Designations of integrated components are concatenated in alphabetical order, or comma-separated if subdivided further (e.g., Aa, Ab). Undesignated components from WDS are indicated by an asterisk (*).

The *Classes* column provides entire sets of comma-separated SIMBAD classifications (e.g., variable types, duplicity, and other stellar characteristics.) in their abbreviated formats for every star. A key to abbreviations can be found at http://simbad.u-strasbg.fr/simbad/sim-display?data=otypes

The *Groups* column flags each subject star as a member of one or more star clusters or stellar associations. Integer values are given here for the number of memberships assigned; the groups are named verbosely in the *GrpNames* column of *biblio.dat*. We tested the membership lists of van Leeuwen (2009 and personal correspondence), de Zeeuw et al. (1999), Zuckerman & Song (2004) and all groups listed in the *New catalog of optically visible open clusters and candidates (V3.0)* ("B/ocl", Dias, et al., 2002-2010) or in Karchenko (2005a & 2005b) with distances less than 800 pc in either catalogue, at which point the groups have become too distant to contain many Hipparcos stars, and parallax distances have become meaningless. We found evidence in the Hipparcos database for 87 groups (it does not necessarily follow from absence of evidence that the group does not exist, since it may contain less bright stars). 42 groups show sufficient separation from the surrounding star field to be classed as probable clusters. Our method of assigning memberships is explained in detail in (§3).

## 2.2   Astrometry (*main.dat*)

Hipparcos astrometry is imported from the *New Reduction* version available at CDS (I/311) since 15 Sept. 2008, in which minor corrections over previous releases were made. While important improvements in parallaxes were realized in HIP2 over HIP1, some HIP2 parallaxes are problematic. Stars with multiple components were solved individually rather than as systems for the sake of expediency (van Leeuwen, 2007) and their astrometric solutions may be grossly erroneous. We compared parallaxes given in HIP2 versus HIP1 and confirmed that the largest parallax discrepancies between reductions overwhelming apply to multiple star systems. We elected to revert to HIP1 astrometry ($r\_HIP$=1) in cases where multiplicity is indicated in *Comp* and the formal parallax error in HIP2 is higher than in HIP1. These criteria hold true for 1922 cases (1.6% of the catalog); otherwise we use HIP2 ($r\_HIP$=2). We validated each criterion individually and note that the mean average of differences in position between reductions for the stars with discarded HIP2 astrometry was 13.66 mas (compared to a 1.68 mas mean average of differences for all the stars in the catalog) and the mean average of differences in proper motion for stars with discarded HIP2 astrometry was 11.38 mas/yr (compared to a 1.52 mas/yr mean average of differences for all the stars in the catalog).

We also consulted the *Tycho-2 Catalogue* (Høg et al., 2000) in preparing our list of proper motions (*pmRA*, *pmDE*). We use Hipparcos proper motions ($r\_pm$=1) where Tycho-2 proper motions are not available. We discard the Hipparcos proper motions and opt instead for Tycho-2 ($r\_pm$=2) if the Hipparcos proper motions go beyond the Tycho-2 error bounds when multiplicity is indicated in *Comp*. This serves to manage short-period binaries adversely affecting Hipparcos' short-epoch measurements. If multiplicity is not indicated, we opt for Tycho-2 when Hipparcos proper motions go beyond the Tycho-2 error bounds by a factor of 3. In all other cases ($r\_pm$=3) we used the mean HIP2 and Tycho-2 measurements weighted by the inverse error squared to derive proper motion values with improved error bounds; this latter, optimal, treatment applies to 92 269 cases (78% of the catalog).

Positions and proper motions are rigorously transformed to the Galactic system (*GLON, GLAT, pmGLON, pmGLAT*) using the definitions for the North Galactic Pole and origin of Galactic longitude ($\alpha_G$= 192.85948; $\delta_G$= +27.1282; $l_\Omega$ = 32.93192) – as adopted in §1.5 of the Hipparcos manual (ESA, 1997). Galactic longitudes and latitudes are rendered in *main.dat* to 8 decimal places so that they can be fully utilized as astrometric coordinates.

Because parallax distance is proportional to the reciprocal of parallax angle, expected stellar distances, after taking account the error distribution, are less than $1000/\pi$. This is a part, but not the main part, of the Lutz-Kelker bias (Lutz et al., 1973, 1974, 1975). We used a numerical integration assuming normally distributed parallax errors and found that when stated parallax errors are less than 20% of parallax values (*e_Plx / Plx* < 0.20), the expected heliocentric distance is given by:

$$R \approx \frac{1000}{\pi} \div \left( 1 + 1.2 \left(\frac{\sigma}{\pi}\right)^2 \right). \qquad\qquad 1$$

The *Dist* column gives distances of stars based either on good parallax data (59 563 stars) or on cluster membership (1 314 stars – see §3). The *e%_Dist* column gives error bounds as a percentage-figure for distances based on individual parallaxes or is null when the *Dist* column value has been fitted to a cluster.

For the 60 877 stars having assigned distances, transverse velocities are listed in column *vT*. Heliocentric distances are also broken down into three-dimensional Cartesian axes (columns *X, Y, Z*), in the direction of the Galactic center, Galactic rotation and the Galactic North Pole.

On the assumption that Sgr A* is stationary at the Galactic barycentre, and using an adopted solar orbital velocity of 225 km s[-1], the proper motion of Sgr A* determined by Reid and Brunthaller (2004) implies a distance to the Galactic centre of $R_0$ = 7.4 ± 0.04 kpc, consistent with recent determinations (Reid, 1993; Nishiyama et al., 2006; Bica et al., 2006; Eisenhauer et al., 2005; Layden et al., 1996). The location of Sgr A* is taken to be (*X, Y, Z*) = (7 400, − 7.2, − 6.0) pc.

| Rank | Source | CDS Catalog ID | # used |
|---|---|---|---|
| 1 | *General Catalogue of Stellar Spectral Classifications, version 2010-Mar* (Skiff, 2010) | B/mk | 41 980 |
| 1 | *Michigan Catalogue for the HD Stars, vols. 1 - 5* (Houk & Cowly., 1975; Houk, 1978, 1982, 1988, 1999) | III/31B, III/51B, III/80, III/133, III/214 | 35 948 |
| 1 | *Catalogue of selected spectral types in the MK system* (Jaschek, 1978) | III/42 | 2 166 |
| 1 | *Search for Associations Containing Young stars* (Torres et al., 2006) | J/A+A/460/695 | 531 |
| 2 | *Hipparcos Input Catalogue, Version 2* (Turon et al., 1993) | I/196 | 34 403 |
| 3 | *The Tycho-2 Spectral Type Catalog* (Wright et al., 2003) | III/231 | 727 |
| 4 | SIMBAD (queried March, 2010) | … | 341 |
| | | | 116 096 |

**Table 2**: Sources of spectral classifications ranked by preference, with number of records drawn from each source, totaling 116 096 Hipparcos stars. Records from the set of first-priority sources are drawn according to publication dates of the primary observations. BibCodes for individual classifications are provided in the XHIP file *biblio.dat*.

## 2.3 Spectrography (*main.dat*)

### 2.3.1 Spectral Classifications

We consulted several sources for spectral classifications, listed in Table 2 by order of preference. We also gave individual preference to entries that contained luminosity classes.

We controlled for erroneous identifications by comparing temperature classes in the B0 to M7 range against Hipparcos *B-V* photometry. Whenever we found a type to be 1 class earlier or 2.5 classes later than a star's photometric color would typify (e.g., if the *B-V* index typifies an F0 star, but the putative type is earlier than A0 or later than K5), we gave preference to alternate sources which type the star in the expected range (when available). A total of 116 096 spectral classifications were matched to Hipparcos stars – over 98% of the catalog – listed in column *SpType*.

| Type | Code | | Type | Code |
|------|------|---|------|------|
| O | 10 | \| | L | 80 |
| B | 20 | \| | T | 90 |
| A | 30 | \| | S | 100 |
| F | 40 | \| | C | 110 |
| G | 50 | \| | R | 120 |
| K | 60 | \| | N | 130 |
| M | 70 | \| | | |

**Table 3:** Machine-sortable numeric codes assigned to spectral classifications listed in the *Tc* column. Sub-classes 0-9 are combined by summation (e.g., "B5" = 25). Luminosity classes (I-VI), converted to numeric integers, are listed in the *Lc* column.

We parsed the spectral classifications into numerical formats as described in Table 3. These are listed in separate columns (*Tc* and *Lc*) in *main.dat* for temperature and luminosity classes respectively. 114 315 spectral types were thus converted; 73 842 of these also have luminosity classes.

### 2.3.2 Radial Velocities

We used 47 major sources of radial velocities published since 1992 (Table 4). A total of 46 392 velocities were assigned to individual Hipparcos stars, listed in column *RV* and errors (when available) in *e_RV*. Although the measured radial velocity of many stars on the main sequence is almost constantly within the bounds of measurement errors, a number of physical stellar characteristics, including duplicity, pulsation, rapid rotation and convection, may contribute larger uncertainties to radial velocities than those imposed by the limits of measurement precision. Gravitational redshift contributes a systematic error, dependent on stellar radius and mass, and is not calculated for individual stars. Zero-point discrepancies and systematic dependencies on velocity, color, and equatorial coordinates have been discovered in large surveys (c.f., Gontcharov, 2006, *Pulkovo Compilation of Radial Velocities* "PCRV"). These are corrected in the PCRV, which is used in preference to the uncorrected data.

Radial velocity data are graded by quality in the *q_rv* column of *main.dat*. An "A" rating signifies data for which we consider that the errors are generally reliable. Sources rated "B" may also contain small uncorrected systematic errors, but still correspond well with our most reliable sources (e.g., stars in the *Geneva Copenhagen Survey* excluded in PCRV). "C"-rated sources tend to correspond less well and may contain larger systematic errors (e.g., RAVE; c.f., Gontcharov, 2007), but their data may still be suitable for analyses of high-velocity populations such as thick-disk and halo stars. "D" ratings signify more serious problems, suggesting that these stars may not be suitable for use in statistical analyses. A "D" rating is assigned whenever the error bounds are not given (indicated by *e_RV* = 999), the star is an unsolved binary, or the star is a Wolf-Rayet or white dwarf star that is not a component of a solved binary, or when different measurements in the "A" sources have given inconsistent results and there is insufficient information to reject a particular result.

| Source | CDS Catalog ID |
|---|---|
| *Thirty New Low-mass Spectroscopic Binaries*, Shkolnik et al., 2010 | … |
| *The Ninth Catalogue of Spectroscopic Binary Orbits,*<br>*version 2009-Aug* ("B/sb9", Pourbaix et al., 2004-2009) | B/sb9 |
| *Kinematics of W UMa-type binaries* Bilir et al., 2005 | … |
| *Kinematics of chromospherically active binaries*, Karatas et al., 2004 | J/MNRAS/349/1069 |
| *Wide binary systems and the nature of high-velocity white dwarfs*, Silvestri et al., 2002, | J/AJ/124/1118 |
| *A spectroscopy study of nearby late-type stars*, Maldonado et al., 2010 | J/A+A/521/A12 |
| *Spectroscopic binaries among Hipparcos M giants*, Famaey et al., 2009 | J/A+A/498/627 |
| *Red giants in open clusters, XIV,* Mermilliod et al., 2008 | J/A+A/485/303 |
| *Local Kinematics of K and M Giants from CORAVEL/Hipparcos/*<br>*Tycho-2 Data,* Famaey et al., 2005 | J/A+A/430/165 |
| *A survey of proper-motion stars. XVI (Table 3),* Latham et al., 2002. | J/AJ/124/1144 |
| *Two distinct halo populations in the solar neighborhood*, Nissen & Schuster, 2010. | J/A+A/511/L10 |
| *Spectroscopic properties of cool stars,* Valenti & Fischer 2005. | J/ApJS/159/141 |
| *2086 Nearby FGKM Stars and 127 Standards*, Chubak et al., 2011. | … |
| *The HARPS search for southern extrasolar planets. XXV,* Santos, 2011.<br>[and previous papers from this series.] | J/A+A/526/A112 |
| *Catalogue of radial velocities of Nearby Stars,* Tokovinin, 1992. | III/191 |
| *Radial Velocities for 889 late-type stars,* Nidever et al., 2002. | J/ApJS/141/503 |
| *Vertical distribution of Galactic disk stars. IV,* Soubiran et al., 2008. | J/A+A/480/91 |
| *Pulkovo radial velocities for 35 493 HIP stars,* Gontcharov, 2006. | III/252 |

Grade A velocities:   35 932
Grade B velocities:    4 239
Grade C velocities:    3 465
Grade D velocities:    <u>2 756</u>
       Total:   **<u>46 392</u>**

**Table 4:** List of "A"-grade radial velocity sources. Our "B"-grade sources, in order of preference are: Lopez-Santiago et al., 2010; Massarotti et al., 2008; Guillout et al., 2009; Soubiran et al., 2003; Holmberg et al., 2007; Gizis et al., 2002; White et al., 2007; Torres et al., 2006; Kharchenko et al., 2007. Our preferred "C"-grade sources are: Montes et al., 2001; Garcia-Sanchez, 2001; Turon et al., 1993; Barbier-Brossat et al., 1994. The unranked C-grade sources are: Saguner et al., 2011; Valentini & Munari, 2010; Boyajian et al., 2007; Griffin, 2006; Beers et al., 2000; Grenier et al., 1999; Chiba & Yoshii, 1998; Fehrenbach et al., 1997; Hawley et al., 1996; Levato et al., 1996; Fehrenbach et al., 1996; Duflot et al., 1995; Reid et al., 1995; Turon et al., 1993; Fehrenbach et al., 1992; Duflot et al., 1992; RAVE DR3 (Siebert et al., 2011) and SIMBAD. We also consulted the *Bibliographic Catalogue of Stellar Radial Velocities* (Malaroda et al., 2010), which does not list error figures (hence $q\_RV$ = D for data acquired indirectly through this source); however, this catalog also helped us discover some of our other sources from which we directly drew complete information.

For rapidly orbiting binaries and other stars showing substantial variations in radial velocity, it is necessary to take an average velocity from multiple observations. If the star is a binary, it may be possible to fit the velocity curve to the orbit, with a further improvement in accuracy.

This has been done for a small number of stars in "A" graded sources, but in most cases the sources give an average. For multiple-observation surveys, errors are calculated from the measurement dispersion, and depend mainly on the properties of the star. These are not commensurate with measurement errors given in single observation surveys. We therefore split the "A" graded sources into multiple- and single-observation surveys, and took a weighted average of each. For stars with measurements in both groups, priority was given to multiple observation surveys, then to B/sb9, and last to measurements from single observation surveys. Because sources rated below "A" may contain systematic errors, they are unsuitable for use in a weighted mean. They are ranked according to broad considerations: general agreement with our most reliable sources, publication date, and best stated error bounds.

Keycodes for radial velocity sources are given in the *r_RV* column of *biblio.dat.* For entries we treated by averaging, the input sources are listed in a comma-separated format.

### 2.3.3 Iron Abundances

We assigned a total of 19 097 iron abundances to individual Hipparcos stars, in column *[Fe/H],* and errors in *e_[Fe/H]* (set to "9.99" if unassigned or unknown). Data quality is rated in *q_[Fe/H];* "A" or "B" ratings are assigned to 18 549 calibrated values; "C" ratings are assigned to the remaining 548 uncalibrated values. A detailed account of our sources and calibration procedure is given in §5.

## 2.4   Stellar Ages (*main.dat*)

Stellar ages and confidence limits are given in the *age, clage,* and *chage* columns of *main.dat*. They are directly imported from GCS3, which were revised to take account of the improved parallaxes published in HIP2.

## 2.5   Stellar Motions (*main.dat*)

Of the 60 877 stars with distances provided the *Dist* column (§2.2), 32 958 are associated with radial velocity data and hence considered "kinematically complete." For these stars, we have computed their approximate motions in three-dimensional space.

### 2.5.1   Linear Approximations

We computed space velocity components – corresponding to the conventional *X, Y, & Z* directions in positional space – in columns *U*, *V*, & *W* respectively. Total heliocentric space velocities are listed in column *UVW*.

Using these straight-line motions, we project distances and timings of solar encounters in column *Dmin* and *Tmin.* Linear trajectories may be considered safe approximations to stellar motions to at least within ±2 Myr (Mülläri & Orlov, 1996). Encounters projected beyond ±10 Myr and white dwarf stars with D-grade radial velocities are omitted, leaving 31 332 included. Gliese 710 (HIP 89825) remains the closest known encounter with reliable data (at 0.2pc, ~1.4 Myr. from now), as similarly reported by Garcia-Sanchez et al., (2001), Dybczyński (2006), and Bobylev (2010). Encounters within 2pc and ±2 Myr not reported by these three sources are summarized in Table 5.

| HIP | Comp. | D_min (pc) | T_min (kyr) | Sp. Type | Dist (pc) | RV km s⁻¹ | e_RV | RV Source |
|---|---|---|---|---|---|---|---|---|
| 32475 | AB | 1.65 | -1525 | F0 IV | 68.59 | 43.9 | 0.6 | (Gontcharov, 2006) |
| 19946 | | 1.85 | -532 | G0 | 79.10 | 145.4 | 0.1 | (Chubak, 2011) |
| 21539 | | 1.91 | -137 | K5 V | 34.96 | 248.0 | N/A | (Barbier-Brossat, et al., 1994) |
| 34617 | ABC | 1.93 | -1220 | F4 V | 42.06 | 33.6 | 0.6 | (Gontcharov, 2006) |

**Table 5:** Solar encounters within 2pc and 2Myr stated in *main.dat*, but not reported in Garcia-Sanchez, et al. (2001), who tabulated approaches nearer than 5pc for 155 stars, Dybczyński (2006), who tabulated approaches nearer than ~2.5pc for 46 stars, or Bobylev (2010), who tabulated approaches nearer than ~2pc for 14 stars currently within ~30pc. Minimum distance (Dmin) is given in parsecs; time of minimum distance (Tmin) is given in kyr ago (when negative) or kyr hence (when positive). Current distances (Dist), radial velocities (RV), and formal radial velocity errors (e_RV) are also shown. White dwarf stars which have spectroscopic radial velocities largely affected by gravitational redshifts are not considered.

### 2.5.2 Elliptical Approximation

Stellar orbits are not strictly elliptical but can usefully be regarded as precessing ellipses. We derive elliptical approximations by adopting the location of the Galactic center specified in §2.3, a total solar velocity in the direction of Galactic rotation of 225 km s⁻¹, and LSR values $(U_0, V_0, W_0)$ as $(-14.0, -14.5, -6.9)$, km s⁻¹. These values for the LSR were redetermined using the preferred methods of Francis & Anderson (2009a) but with an improved sample derived from XHIP itself (a full report is the subject of a work in preparation).

We give calculations for orbital eccentricity in column *ecc*. Column *phi* gives the angle, in degrees, subtended at the Galactic center by the star and the projected position of the pericenter of its orbit. Positive angles place the star's pericenter in the direction of Galactic rotation – negative angles in the direction of Galactic anti-rotation. Columns *a, b, c,* & *L* give the semi-major axis, semi-minor axis, focus-to-center distance, and the semilatus rectum of the ellipse. *Rmin* and *Rmax* give the minimum (pericenter) and maximum (apocenter) distances to the Galactic center; orbits with low to moderate eccentricities yield good estimates of these distances.

The parameters for the Solar orbit in this regime (not included in the XHIP database) are: *ecc* = 0.16, *phi* = 26.0, *a* = 8 685, *b* = 8 573, *c* = 1 388, *L* = 8 464, *Rmin* = 7 297, *Rmax* = 10 074.

## 2.6 Exoplanets and Circumstellar Disks (*main.dat*)

We queried the *The Extrasolar Planets Encyclopaedia* (Schneider, 2012) to identify exoplanet host stars. 364 Hipparcos stars hosting 465 planets were found. We also queried the Catalog of Resolved Circumstellar Disks (McCabe, Stapelfeldt & Pham, 2011), finding 40 matches to Hipparcos stars.

The *Planets* column gives a positive integer for the number of exoplanets discovered per star, a zero value if the star only has a circumstellar disk, or is null if neither. The *Methods* column lists the means of exoplanet discovery categorized in the Encyclopaedia (RA = "radial velocity or astrometric methods"; I = "imaging"; T = "timing"; X= "transit"). When multiple methods of detection apply, these designations are comma-separated.

## 2.7 Photometry (*photo.dat*)

### 2.7.1 Presentation

Hipparcos broadband magnitudes, their associated flags, and various Johnson passband magnitudes are arranged in ascending order of effective wavelength, as listed in Table 1. For the 60 770 stars with distances provided the *Dist* column in *main.dat* (see §2.2), absolute magnitudes in each band are computed. Luminosities are also computed based on V with a bolometric adjustment applied from correction figures supplied by Masana et al. (2006) when available, otherwise (when B-V < 1.6) from the B-V indexed tables of Flower (1996). No modeling of interstellar absorption is factored into these calculations and should be used with caution for stars at large distances and/or low Galactic latitudes. They may also reflect integrations of multiple components (see *m_Hpmag*).

For the 31 295 stars with projected solar encounters in *main.dat* (see §2.5), the *magmin* column gives the estimated magnitude at closest approach, calculated from the absolute magnitude *V_Mag* at the projected distance of *Dmin (main.dat)*. Neither considerations of stellar evolution nor changes in interstellar absorption are factored into the *magmin* estimates. Theta Columbae (HIP 29034) has the brightest projected magnitude of all the close-encounter candidates: -5.10 (at 2.11pc, ~4.8 Myr ago). The Canis Major hot stars Adhara and Mirzam (HIP 33579 & 30324) are the two next brightest at magnitudes -4.12 and -3.78 (at 9.35 and 10.61 pc, ~4.4 Myr ago).

### 2.7.2 Sources

Broadband (*Hpmag*) magnitudes, their associated flags, *V* magnitudes, and color indexes were imported directly from the Hipparcos Catalog. *B* and *I* magnitudes were extrapolated from Hipparcos' *B-V* and *V-I* index, when available, otherwise they are sourced from the Hipparcos BTmag column (Tycho Magnitude System). *U* and *R* magnitudes were retrieved from SIMBAD.

*J, H,* and *K* magnitudes come from 2MASS All-Sky Catalog of Point Sources (Cutri et al., 2003). 2MASS cross-identifications were selected from either SIMBAD, Reed (2007), or our own VizieR search. The high star density of 2MASS, especially in the Galactic plane, imposes challenges to proper identifications. As a measure of quality assurance, we rejected cross-identifications in which *V* magnitudes of Hipparcos stars were brighter than putative 2MASS *J* magnitudes or when multiple plausible matches were found. A total of 114 738 2MASS matches (97% of the catalog) were assigned.

## 2.8 Bibliography (*biblio.dat*)

Constellation membership is provided in the *Con* column. We queried Roman (1987) using the VizieR service (CDS Catalog VI/42) to obtain constellation placement. Standard 3-letter abbreviations are used. The *Atlas* column lists the page number on which the star is best-placed in the Millennium Star Atlas (Sinnott & Perryman, 1997). The adjacent *Coords* column gives equatorial coordinates in a compact sexagesimal format.

The *Name* column lists "classic" star identifications that are most suitable for communicating star identities to the public. We consulted Kostjuk (2002) to obtain cross-identifications with Bayer-Flamsteed designations and proper star names. We also queried SIMBAD to obtain common names and designations in the Gliese, Ross, Wolf, Lalande, and HR catalogs. The *Name* column lists up to two unique identifications, prioritized in the order just described.

The *GrpNames* column lists star membership assignments to clusters and associations. Where overlapping memberships are assigned, the names are comma-separated.

Sources for individual component designations, spectral classifications, radial velocities, and iron abundances are listed in the *r_Comp*, *r_SpType*, *r_RV*, and *r_[Fe/H]* columns respectively. Standard SIMBAD/NED BibCodes are supplied for known sources. Occasionally, spectral classifications and radial velocities obtained through SIMBAD are unreferenced; these instances are identified with the pseudo-code "2010…SIMBAD". The BibCodes "1993BICDS..43....5T" (Hipparcos Input Catalog) and "2003AJ....125..359W" (Tycho-2 Spectral Type Catalog) in the *r_SpType* column of *biblio.dat* are further augmented (after a colon) with the content of the r_Sp or r_SpType columns from those catalogs respectively. Sources for iron abundances, in *r_[Fe/H]*, are indexed with comma-separated numeric codes which correspond to the BibCodes in *keycodes.dat*.

## 3  Clusters and Associations

We define moving groups as stars sharing a common motion and localized in a region of space. They are distinguished from streams, which are all-sky motions, which are parts of the spiral structure of the Galaxy, as shown by Francis & Anderson (2009b). Moving groups will be termed 'clusters' if they are gravitationally bound, and 'associations' otherwise. Associations typically consist of young stars originating in the same process, resulting from the collisions between outward bound gas clouds (corresponding to the Hyades stream) and clouds following the spiral arm.

Because of the likelihood of chance alignments it is important to calculate group memberships on the basis of complete kinematic data, and with the most accurate and extensive information available. Since the purpose of XHIP is to provide this information, it is appropriate to recalculate group memberships for all clusters and associations containing a reasonable number of Hipparcos stars with known radial velocities. It is also necessary to identify group memberships to eliminate any selection bias arising from the fact that certain radial velocity surveys have studied moving groups, and to avoid the accidental weighting of statistical properties of a population towards properties of groups simply because of the numbers of stars they contain.

To identify the membership of a moving group it is necessary to match both the position and the 3-velocity of each star to within a region of a six-dimensional, position × velocity, space. Identifications can only be given with certainty for stars with known radial velocities, and accurate parallaxes. In practice, many stars do not have known radial velocities and, even using HIP2, parallax errors lead to distance errors much greater than the size of the group. Consequently, there remains some uncertainty in group memberships. We determined group memberships iteratively, by testing each star in XHIP under the condition:

$$\sum_x \frac{(x - \bar{x})^2}{\sigma_x^2} < n^2 \qquad\qquad 2$$

where, for each candidate star, *x* runs over the dynamical variables, *Plx*, *RA*, *DE*, *RV*, *pmRA*, *pmDE*, except for groups occupying a large region of the sky, when *x* runs over the Cartesian variables *X, Y, Z, U, V, W*. $\bar{x}$ and $\sigma_x$ are the mean and standard deviation of the group in the previous iteration. *n* is the largest integral or half integral value for which a stable group was found under the iteration. Cartesian variables were used for the Hyades cluster, the Ursa Major association, the AB Doradus and β Pictoris moving groups and the Lower Centaurus Crux, Upper Centaurus Lupus, Upper Scorpius and Tucana/Horologium associations. It is necessary to use Cartesian variable when the angular size of the cluster substantially affects radial velocity and proper motions, but it is better to use radial variables where possible because radial distance errors are much greater than angular errors, and hence the error ellipsoid is aligned with the radial direction.

Eq. 2 can be justified from statistical considerations, and, in the case of clusters, dynamically from the virial theorem. Broadly, for a gravitationally bound group, the further a star is from the centre of the group the nearer its velocity must be to that of the centre of gravity of the cluster. Associations are not gravitationally bound, and do not have the ellipsoidal shape suggested by eq. 2. Memberships of associations are necessarily less accurate than those of clusters, but eq. 2 still has merit, by providing an objective criterion for the existence of a group, and an initial list of candidate stars.

We initialized the iteration using a list of group stars taken from van Leeuwen (2009 and personal correspondence), de Zeeuw et al. (1999), Zuckerman & Song (2004), or group coordinates given in B/ocl or Karchenko (2005a & 2005b)  together with typical group dimensions. The cluster was considered "found" when the group became stable under iteration of eq. 2. For associations, the found groups are not necessarily those intended by those who originally named them. In some cases convergence has been found for larger groups than those originally intended, with the result that our lists contain overlaps.  For example, the Orion Molecular Cloud Complex contains a number of groups including the Orion Nebula (M42), The Horsehead Nebula, Collinder 70, NGC 1980, NGC 1981, and the Running Man Nebula (NGC 1977) which have similar motions and which can be regarded as parts of the larger group.

$n$ is a measure of the concentration of a cluster, and of its separation from the surrounding star field, and (because it is rigorously defined) may be preferred to the Trumpler classifications I – IV. Since they are gravitationally bound, clusters are more compact than associations. As a result the largest value of $n$ giving convergence can be expected to be greater for a cluster than for an association. In practice, we found that a maximum value of $n = 3.5$ is typical for an association. For clusters $n \geq 5$ is usual. The lower value of $n$ for associations shows poor dynamical separation from surrounding stars, indicating that the majority of associations are just randomly dense regions arising in much greater processes in which stars are formed. This was also shown by the fact that a number of associations found in the Pleiades stream have overlapping memberships. We rejected groups with $n < 3.5$ or with fewer than four Hipparcos stars with known radial velocities.

For most well-known clusters our 6-dimensional fitting procedure leads to a larger number of candidates and/or smaller dispersion than is found in lists obtained from the literature. For groups using spherical coordinates, we extended the list of candidate stars to include stars without known radial velocities by reducing the right-hand side of eq. 2 so as to leave the standard deviation of each dynamical variable similar to its value for the core group of stars with complete dynamical information. This method does not apply to groups found using Cartesian coordinates.

## 4    Test for Kinetic Bias

XHIP compiles radial velocities from a variety of sources. Binney et al. (1997) have claimed that such compilations should not be used in kinematic studies because they contain a bias toward high proper motion stars. They did not give a statistical analysis for their conclusion, but justified it from a graph (their fig. 2) with a logarithmic scale which exaggerates evidence of bias by two orders of magnitude. In fact a bias towards high proper motion will result if, as may be expected, near stars are chosen in surveys, but bias in proper motion is not, in itself, evidence of bias in velocity.

After restricting the sample to stars with radial velocities available to Binney et al. it remains impossible to reproduce the level of bias which they reported. The reason is that they limited their sample by luminosity distance. In view that Hipparcos provides accurate parallax distances for nearby stars, this decision appears a little strange. Since they did not give details of their distance model, it is not possible to reproduce either their data or their result. By comparison with the (rather limited) population of stars with radial velocities available at the time, with or without a cut on parallax distance, it is possible to determine that the selection bias reported by Binney et al. was introduced by the authors themselves through their choice of distance model.

The number of stars with available radial velocities is now hugely greater, and it is important to assess the population for signs of bias. It is to be expected that XHIP will contain a selection bias towards radial velocities for stars in clusters, because these have been the subject of a number of surveys. Irrespective of such a selection bias, in a kinematic study of the Galaxy it is desirable to treat moving groups as single objects, since otherwise they will weight statistical properties incorrectly. In view of the large population in XHIP, it is sufficient to simply remove clusters and associations, since individual objects will have negligible effect on the statistical properties of the population as a whole.

To test for a selection bias in the remaining population we first removed 4 426 stars identified as either belonging to clusters and associations (§3), or which are identified as non-primary stellar components (§2.1). We removed stars with zero or negative parallaxes, for which no distance or transverse velocity can be calculated. We binned the remaining 109 887 stars by transverse velocity, and plotted the ratio of the number for which we have radial velocities with $e\_RV < 5$ and $q\_RV \neq "D"$ to the number of stars in each bin for the whole population (Figure 1). Plots are also shown for stars in GCS and Famaey, which are deemed to be free from kinematic selection bias. In practice, we found that the exclusion of stars in clusters and associations made almost no visible difference to the plot. For this exercise, using the entire population, transverse velocity was calculated using distance = $1000/Plx$, since the distance correction (eq. 1) is not valid for stars with large proportionate parallax errors.

Figure 1 shows that there is clearly no selection bias toward high transverse velocity stars. A perfectly flat line is not expected because a magnitude-limited sample will have dependencies on stellar type. In turn this will create kinematic dependencies because the velocity distribution in the solar neighbourhood is dependent on factors like temperature and age. Although GCS and Famaey are each chosen to be kinematically unbiased within a range of stellar types, they are not a kinematically unbiased sample of the total population because of the dependency of the velocity distribution on stellar type.

When the population is restricted to 55 889 stars inside 300 pc (calculated using the distance correction, eq. 1), and with parallax errors less than 20%, it is found that the proportion of high velocity stars with known radial velocities appears to rise (Figure 2). However, it is not possible to say that this is due to selection bias, e.g. from radial velocity surveys of low metallicity or halo stars. Even if it is selection bias, the number of stars involved is very small. These stars are removed from statistical analysis by the standard practice of removing outliers. Because of the structure of the velocity distribution, and because the process of removing outliers to a given number of standard deviations of the mean velocity does not converge, Francis & Anderson (2009a) removed stars outside of a nominal "5σ" velocity ellipsoid:

$$\frac{(U+12)^2}{200^2} + \frac{(V+42)^2}{200^2} + \frac{(W+7)^2}{120^2} < 1. \qquad 3$$

This corresponds approximately to a 4 s.d. cut on each axis for the population of old stars, and to over 6 s.d. for the remaining population. This cut removes only 109 stars from a total population of 26 114 stars with radial velocities, and only 39 from 26 038 having transverse velocities less than 200 km s$^{-1}$, but substantially alters the proportions of stars with radial velocities for transverse velocities greater than 150 km s$^{-1}$.

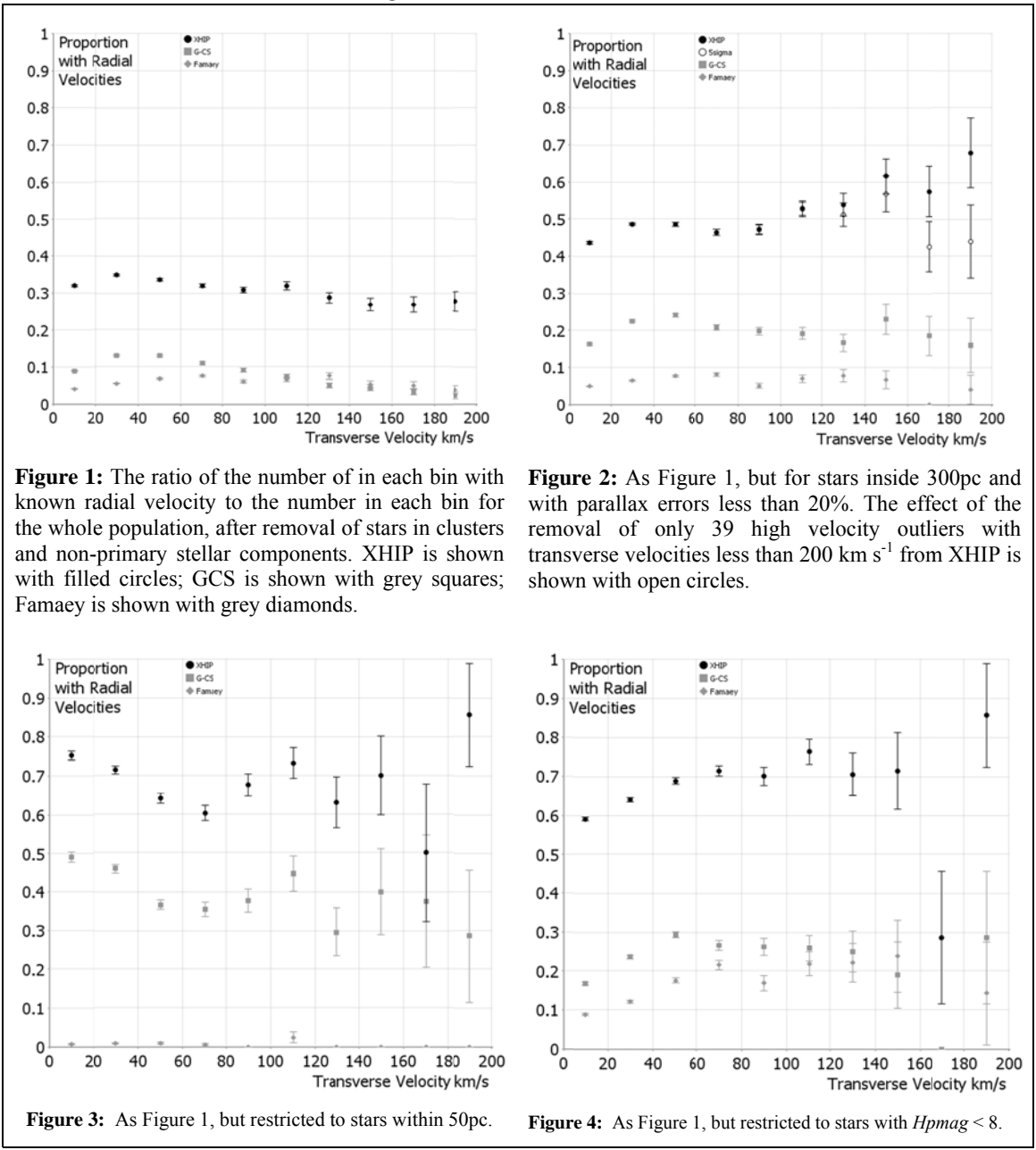

**Figure 1:** The ratio of the number of in each bin with known radial velocity to the number in each bin for the whole population, after removal of stars in clusters and non-primary stellar components. XHIP is shown with filled circles; GCS is shown with grey squares; Famaey is shown with grey diamonds.

**Figure 2:** As Figure 1, but for stars inside 300pc and with parallax errors less than 20%. The effect of the removal of only 39 high velocity outliers with transverse velocities less than 200 km s$^{-1}$ from XHIP is shown with open circles.

**Figure 3:** As Figure 1, but restricted to stars within 50pc.

**Figure 4:** As Figure 1, but restricted to stars with *Hpmag* < 8.

We can conclude that the possible inclusion in XHIP of a small excess number of high velocity stars within 300 pc with known radial velocities is of no importance to statistical analysis of the data as a whole.

The low proportion of stars with radial velocities in the first bin is related to the fact that Hipparcos is magnitude limited. The effect can be reversed by imposing a 50 pc distance bound (Figure 3), or increased by putting a stricter magnitude bound on the population (Figure 4). The shape of the plot under different constraints has to do with the structure of the velocity distribution, together with the fact that the population of stars with known radial velocities is dominated by GCS, which is principally a population of F & G dwarfs and includes some K & M dwarfs.

Because the large radial velocity surveys, Famaey and GCS, have selected particular stellar types, because Hipparcos is magnitude limited, and because of the dependency of the velocity distribution on stellar type, it is impossible to avoid a bias in any stellar population for which we have complete kinematic data. The only way to completely avoid bias is to obtain complete kinematic information on all stars within a neighbourhood. This will become available with Gaia. In practice, even if complete data were available, because of the structure of the velocity distribution, standard statistical measures of the population are of limited value, and cannot be used, for example, to calculate the LSR (Francis & Anderson 2009a). It is therefore vital for dynamic and kinematic studies of the Galaxy to make use of all the available data, which we have collated into XHIP, and to analyse this data in an appropriate matter, rather than by following routine procedures.

## 5   Metallicity Calibration

Metallicity, normally measured as iron abundance [Fe/H], is one of the most important parameters in stellar evolutionary theory. Regrettably, the value of figures from the literature is limited because data is not homogeneous. Measurements have been taken using many spectrometers, at differing resolutions, and procedures are not standardized between observation programs. For example, different observers may base figures on different spectral lines. In many instances errors are not given, and in some cases where margins of error are stated, the correlation with other data shows that they are unrealistic. We therefore elected not to use stated errors, but to assess the accuracy of each database from our own analysis.

*The PASTEL catalogue of stellar atmospheric parameters* ("PASTEL," Soubiran et al., 2010) is a compilation containing metallicities from over 450 separate sources. The data is not homogeneous and is of variable quality. For the purpose of calibration we treated each source separately. We also included data from Wu et. al. (2011), Santos et al. (2011), Masseron et al. (2010), Gebran et al. (2010), Schuster (2010), Ramirez et al. (2009), Guillout et al. (2009), Nissen & Jenkins et al. (2008), Soubiran et al. (2008), Robinson et al. (2007), GCS3, Taylor (2005), Soubiran et al. (2003), Beers et al. (2000), and Chiba & Yoshi (1998).  We were not able to calibrate data from Gebran et al., or the majority of sources in PASTEL, either because the source contains data for too few Hipparcos stars, or because it correlates badly with other data, leading us to think it is unreliable. For our final table of iron abundances, we recalibrated the 175 best databases, and then found mean values weighted by quality of the source for 18 549 stars, as determined from its correlation with other data. We describe details of this procedure below.

Our homogenization procedure considers correlations between databases, and adjusts the zero point (or intercept) and the scale (or slope) to achieve a match. We found that variables, peculiar stars and stars with envelope of type CH are likely to produce inconsistent measurements of [Fe/H] between databases. Since outliers have an adverse effect on the accuracy of correlations, these types were removed from the calibration procedure (though not from the databases or from our final table). Metallicities for these types of stars are intrinsically less reliable than others in XHIP, and we do not give errors for them. We also removed from the database about eighty individual measurements which appeared inconsistent with a number of other measurements on the same star.

In order to produce a homogenized database from the many available sources we require a large basis of reliable data from high resolution spectrographs. After comparing correlations between the best available sources, we selected two of the largest and most reliable sets found from high resolution spectrographs, Wu (2011) and Taylor (2005), which contain over 2000 stars between them, and have 241 stars in common after removal of outliers as described above, with a correlation of 0.985. This is high, but it is not the highest pairwise correlation between datasets.

We chose to calibrate to Wu, and adjusted the zero and scale of Taylor accordingly. We split the error equally between the sets, finding 0.044 dex, in good agreement with the claimed errors for these sets. We used the inverse square of this value to weight the measurements of Wu and Taylor in our combined database. We repeated the procedure for each remaining dataset in turn, calibrating each dataset to the combined sample of previously calibrated data. This was done broadly in order of highest correlation first, but there is no rigorous best order, since changing the order slightly changes the correlations. The ordering is not critical, since any difference in the final results due to minor changes of ordering of datasets is much less than the errors. We promoted larger datasets, and demoted those with few overlapping measurements with the combined dataset so far. At each iteration, the combined dataset was extended by using the mean, weighted by half the inverse mean squared difference between the value given in the current dataset and the value calculated in the previous iteration. We eliminated datasets with fewer than five stars in common with the combined set, or for which the correlation with the combined set is less than 0.6, since calibrating these sets is effectively meaningless. Although these bounds are perhaps low, we felt that the high error margins and low weights given for poorly calibrated data justified their inclusion.

Weighted mean metallicities for variables, peculiar stars, and stars with envelope of type CH, have been included with $q\_[Fe/H]$ = "B". In addition, 548 mean values of [Fe/H] from uncalibrated datasets are given with $q\_[Fe/H]$ = "C", giving a total of 19 097 iron abundances.

The major issues regarding the accuracy of metallicity information concern difficulties in the measurements. With typical quoted errors in the best databases of about 0.05 dex, it is impossible to claim great accuracy for any measurement. The standard procedure for finding a best estimate in such a case is to calibrate all measurements to one scale, and to take a mean weighted according to the strength of the correlation. The reliability of the method is seen in the fact that 150 out of our final 175 databases have correlation coefficients to the main body greater than 0.9.

The largest dataset is GCS3, from which metallicities for 13 973 stars are available. Questions are raised about accuracy and correlation because GCS3 gives photometric, not spectrographic, metallicities, and provides no error bounds. GCS3 metallicities are established from a heuristic calibration equation which relates *uvbyβ* photometry to spectroscopic metallicities for particular stars. Clearly this means that the correlation of GCS3 to a given database depends on such factors as the stellar types in that database, and the methodology used to determine metallicity. It follows that for stars of types differing from those used in the calibration, GCS3 may be quite inaccurate. For example, the correlation coefficient between GCS3 and Wu is 0.987, between GCS3 and Taylor it is 0.938, but between GCS3 and Guillout et al. it is only 0.483. Guillout et al. is a survey of young stars, whose spectra may present difficulties due to such factors as high rotational velocity. The correlation between Guillout et al. and the population excluding GCS3 is 0.79, which is relatively low, and it is necessary to calibrate it to the main body before calibrating GCS3, or it would be excluded altogether for having too low a correlation. We therefore chose to place GCS3 last in order of databases calibrated to the main body. When available, we used the recalibration of the GCS3 metallicities by Casagrande et al. (2011), which is better motivated physically, and for which the correlation with other datasets is 0.935, a little better than 0.928 found for the original GCS3 metallicities. After recalibration of the zero and slope, the standard error for Casagrande is 0.08 dex, whereas for GCS3 it is 0.09 dex.

## 6 Conclusion

XHIP is the most comprehensive compilation of supplemental data available for Hipparcos stars to date. In a single database we have provided 45 392 radial velocities and 18 549 mean homogenized iron abundances from 175 sources, together with 548 metallicities from sources which we are not able to calibrate to the main body. Although radial velocity surveys have tended to concentrate on stars in clusters, the number of cluster stars is small compared to the number in the database, and these stars are easily removed. There is no sign of kinematic bias in the remaining database, and certainly no sign of bias toward high velocity stars.

**Data Retrieval**

XHIP can be retrieved from the Centre de Données astronomiques de Strasbourg (CDS Catalog V/137B).

**Acknowledgements**

We are pleased to acknowledge that this work made use of the *Hipparcos* catalog from the European Space Agency (ESA), the SIMBAD database and VizieR service operated at the Centre de Données astronomiques de Strasbourg (CDS), NASA's Astrophysics Data System, the Washington Double Star Catalog maintained at the U.S. Naval Observatory, and data products from the Two Micron All Sky Survey – a joint project of the University of Massachusetts and the Infrared Processing and Analysis Center at the California Institute of Technology and funded by the National Aeronautics and Space Administration and the National Science Foundation. E.A. especially thanks Brian Skiff at Lowell Observatory for his unfailing responses to general inquiries throughout the development of this compilation. François Ochsenbein at CDS also answered numerous inquiries, made useful suggestions, and lent technical support on many occasions. Orlando Hugo Levato at the Instituto de Ciencias Astronómicas, Caroline Soubiran at the Laboratoire d'Astrophysique de Bordeaux, Poul Erik Nissen at the University of Aarhus, David Latham at Harvard University,

Alessandro Massarotti at Stonehill College, Arnaud Siebert at Observatoire de Strasbourg, and Ulisse Munari, Tenay Saguner, and Marica Valentini at the Astronomical Observatory of Padova, and Carly Chubak at UC Berkeley responded to inquiries which facilitated the inclusion of recent radial velocity data that they published (or co-published) into this compilation.